\documentstyle [12pt,aaspp4]{article}
\def\erfc{{\rm {erfc}}}
\def\bx{{\bf x}}
\def\br{{\bf r}}

\def\bv{{\bf v}}
\def\mpc{\,{\rm {Mpc}}}
\def\mpch{\,h^{-1}{\rm {Mpc}}}
\def\kms{\,{\rm {km\, s^{-1}}}}

\def\Deltae{\Delta_E}
\def\deltac{\delta_c}
\def\bxi{{\overline \xi}}
\def\vdisper{{\langle v^2_{12}(r)\rangle}}
\def\vdisperr{{\langle v^2_{12}(r)\rangle}^{1/2}}
\def\vone{{\langle v_1^2\rangle}}

\def\ref{\parskip=0pt\par\noindent\hangindent\parindent
    \parskip =2ex plus .5ex minus .1ex}
\def\gs{\mathrel{\raise1.16pt\hbox{$>$}\kern-7.0pt
\lower3.06pt\hbox{{$\scriptstyle \sim$}}}}
\def\ls{\mathrel{\raise1.16pt\hbox{$<$}\kern-7.0pt
\lower3.06pt\hbox{{$\scriptstyle \sim$}}}}
\def\gtsima{$\; \buildrel > \over \sim \;$}
\def\ltsima{$\; \buildrel < \over \sim \;$}
\def\prosima{$\; \buildrel \propto \over \sim \;$}
\def\gsim{\lower.5ex\hbox{\gtsima}}
\def\lsim{\lower.5ex\hbox{\ltsima}}
\def\simgt{\lower.5ex\hbox{\gtsima}}
\def\simlt{\lower.5ex\hbox{\ltsima}}
\def\simpr{\lower.5ex\hbox{\prosima}}
\def\la{\lsim}

\begin{document}
\title {Analytical Approximations to the Low-Order Statistics 
\\ of Dark Matter Distribution}
\author {H.J. Mo, Y.P. Jing, G. B\"orner} 
\affil {Max-Planck-Institut f\"ur Astrophysik,
Karl-Schwarzschild-Strasse 1, 85748 Garching, Germany}
\slugcomment{submitted to MNRAS}
\received{---------------}
\accepted{---------------}

\begin{abstract}
We show that in a hierarchical clustering model the low-order  
statistics of the density and the peculiar velocity fields
can all be modelled semianalytically for a given cosmology and 
an initial density perturbation power spectrum $P(k)$. 
We present such models
for the two-point correlation function $\xi(r)$, the
amplitude $Q$ of the three-point correlation function, the mean 
pairwise peculiar velocity $\langle v_{12}(r)\rangle $, the pairwise
peculiar velocity dispersion $\langle v_{12}^2(r)\rangle$, and 
the one-point peculiar velocity dispersion $\langle v_1^2 \rangle$.
We test our models against results derived
from N-body simulations. These models 
allow us to understand in detail how these statistics 
depend on $P(k)$ and cosmological parameters.
They can also help to interpret, and maybe correct for,  
sampling effects when these statistics are 
estimated from observations. The dependence of the small-scale
pairwise peculiar velocity dispersion on rich clusters 
in the sample, for instance, can be studied quantitatively.
There are also significant implications for the reconstruction 
of the cosmic density field from measurements in redshift space. 
\end{abstract}


\section {INTRODUCTION}

The large scale structure of the Universe is believed to have developed
from small perturbations (usually assumed to be Gaussian) 
of the matter density field by gravitational instabilities. 
Under these assumptions the clustering pattern and velocity
field observed today are determined by the initial conditions via the 
perturbation power spectrum [$P(k)$] and the cosmological parameters such as  
$\Omega_0$, the cosmic density parameter. It is therefore possible  
to derive constraints on model parameters from the observed density and 
velocity distributions of galaxies. 

There are two fundamental problems 
that must be addressed here:
First, since it
is unlikely that galaxies are completely unbiased testing particles 
of the matter density field, we need to understand any such bias
in order to make meaningful comparisons between models and the 
observations of the galaxy distribution. Second, even if the observed 
galaxy distribution traces the matter distribution, we still need to  
understand how the observed distribution is related to the 
parameters that describe a cosmogonical model. The latter is by no 
means trivial, because the clustering pattern and velocity field 
observed today are nonlinear. N-body simulations are usually 
invoked to find a solution to this problem. However, 
as discussed comprehensively by 
Peebles (1980), a complete statistical description of the 
clustering process is also provided by the whole BBGKY hierarchy for    
the distribution function of mass particles. 
Indeed, adding some assumptions, one can solve the 
low order moments of the distribution function directly from the BBGKY 
equations. Although an incomplete description,
such solutions are extremely useful for us to gain physical 
insights into the clustering process.  These low order moments  
are also the most important ones, not only because they describe the
most fundamental part of the clustering process but also because
in practice they are the ones that can be measured from observations. 

In this paper, we show that semianalytical models can be constructed  
for all low order moments of the distribution function. 
These models allow such moments to be calculated directly from
the initial density perturbation power spectrum for a given   
cosmological model, and thus provide us with a clear picture of
how these moments are related to various model parameters. 
We present our models and test them against
results from N-body simulations in Section 2. In Section 3 we 
demonstrate how our models can be used to help us 
to understand better some sampling effects 
when these moments are to be estimated from observations.
Finally, in Section 4 we discuss the implications of our models
to the reconstruction of real-space quantities from 
redshift distortions, and to the construction of some new
statistics that can discriminate current models of structure 
formation.   

\section {MODELS AND TEST BY N-BODY SIMULATIONS} 

In this paper we will consider cosmogonies in which the Universe is dominated
by cold dark matter (CDM). The cosmology is described by the cosmological
matter density ($\Omega_0$), the cosmological constant ($\lambda_0$) 
and the Hubble constant 
($H_0=100h\kms\mpc^{-1}$). When $\lambda_0\ne 0$, we assume
the universe to be flat so that $\Omega_0+\lambda_0=1$. 

The initial power spectrum is 
[Bardeen et al. (1986)]:
$$
P(k)\propto {kT^2(k)},
\eqno(1a)
$$
with
$$
T(k)={{\rm ln}(1+2.34q)\over 2.34 q}
\left\lbrack 1+3.89 q+(16.1q)^2+(5.46q)^3+(6.71q)^4
\right\rbrack^{-1/4}
\eqno(1b)
$$
and
$$
q\equiv {k\over \Gamma h {\mpc}^{-1}} .
\eqno(1c)
$$
Following Efstathiou et al. (1992), we have
introduced a shape parameter, $\Gamma\equiv \Omega_0 h$,
for the power spectrum. (Note that the $\Gamma$ defined here is
not exactly the same as that of Efstathiou et al.). 
When $\Gamma$ is treated as a free parameter, equation (1) can also be 
used to describe the power spectra in other structure formation
models, such as the
mixed dark matter (MDM) models (see e.g. Ma 1996).  
The {\rm rms} mass fluctuation in top-hat windows with
radius $R$, $\sigma (R)$, is defined by
$$
\sigma^2 (R)=\int_0^\infty {dk\over k}\Delta^2(k) W^2(kR),
\eqno(2a)
$$
where 
$$W(x)={3(\sin x-x\cos x) \over x^3}
\eqno(2b)
$$
is the Fouriour transform of the top-hat window function,
and
$$
\Delta^2(k)=
{1\over 2\pi^2} k^3P(k)
\,.\eqno(3)
$$
For convenience, we will refer to $\Delta^2(k)$ as the
power variance.
We normalize $P(k)$ by specifying $\sigma_8\equiv \sigma(8\mpch)$.

\subsection {N-Body Simulations}   

Before presenting our analytic approximations, we first give a
brief summary of 
the N-body simulations to be used to test the models.

We will use results derived from various ${\rm P}^3{\rm M}$ 
N-body simulations (see Jing et al. 1995 for details
on these simulations). 
Each simulation can be characterized by four model parameters 
($\Omega_0, \lambda_0, \Gamma, \sigma_8$), as discussed above, 
and three simulation parameters: the box size $L$ (in $\mpch$), 
the number of simulation particles $N_p$ and the
effective force resolution $\epsilon$ (in $\mpch$). 
The model and simulation parameters for all simulations are
listed in Table 1. For each simulation we give it a name which 
is listed in the first column of the table and will be used throughout
the paper to refer to the simulation.  

\subsection {The Two-Point Correlation Function}   

The evolved two-point correlation function $\xi (r)$ is related 
to the {\it evolved} power variance $\Deltae^2 (k)$ by
$$
\xi(r)=\int_0^\infty {dk\over k} \Deltae^2(k) {\sin kr\over kr}.
\eqno(4)
$$
Thus, in order to get $\xi(r)$ we need an expression for $\Deltae(k)$.
Following the original argument of Hamilton et al. (1991),
Peacock \& Dodds (1994), Jain, Mo \& White (1995),
Padmanabhan et al. (1996) and Peacock \& 
Dodds (1996, hereafter PD) have obtained fitting formulae which relate 
$\Deltae$ to the initial density spectrum for a given cosmological model. 
The latest version of such a fitting formula is given in PD. 
PD have checked their fitting formula for $\Deltae(k)$
by a set of N-body simulations.
Here we use that formula and provide new tests  
for its predictions for $\xi(r)$ using results from (independent)  
N-body simulations. 

The evolved power variance $\Deltae$ 
is assumed to be a function of its linear counterpart $\Delta$:
$$
\Deltae^2(k_E)=f[\Delta^2(k_L)],
\eqno(5)
$$
where $k_L\equiv [1+\Deltae^2(k_E)]^{-1/3}k_E$. The functional form
of $f$ is approximated by (see PD for details) 
$$
f(x)=x\left\lbrack {1+{\cal R}(x) \over 1+{\cal Q}(x)}\right\rbrack
^{1/\beta},
\eqno(6)
$$
where 
$${\cal R} (x)=B\beta x+(Ax)^{\alpha\beta} ,
\eqno(7a)
$$
$${\cal Q} (x)=A^{\alpha\beta}[g(a)]^{3\beta}
V^{-\beta}x^{(\alpha-1/2)\beta},
\eqno(7b)
$$
with 
$$A=0.482[1+n(k_L)/3]^{-0.947}     ,\eqno(8a)$$
$$B=0.226[1+n(k_L)/3]^{-1.778}     ,\eqno(8b)$$
$$V=11.55[1+n(k_L)/3]^{-0.423}     ,\eqno(8c)$$
$$\alpha=3.310[1+n(k_L)/3]^{-0.244} ,\eqno(8d)$$
$$\beta=0.862[1+n(k_L)/3]^{-0.287}  ,\eqno(8e)$$  
$$n(k_L)\equiv {d\ln P\over d\ln k}(k=k_L/2) .\eqno(8f)$$
The function $g(a)$ in equation (7b) is the linear growth
factor at the cosmic time corresponding to the
expansion factor $a$, and for given $\Omega_0$   
and $\lambda_0$, it is accurately described by
(see Carroll, Press \& Turner 1992):
$$
g(a)={5\over 2}\Omega\left\lbrack \Omega^{4/7}
-\lambda +(1+\Omega/2)(1+\lambda/70)\right\rbrack ^{-1}, 
\eqno(9a)
$$
where 
$$\Omega\equiv \Omega (a)={\Omega_0\over a+\Omega_0(1-a)+\lambda_0(a^3-a)},  
\eqno(9b)
$$
$$
\lambda\equiv \lambda(a)
={a^3\lambda_0\over a+\Omega_0(1-a)+\lambda_0(a^3-a)}.  
\eqno(9c)
$$
For later use we also define 
$$
D(a)\equiv ag(a) .
\eqno(9d)
$$

It is clear that $\Deltae$ at any 
wavenumber is determined by $\Delta$
for a given cosmology, and we can just use equation (4) to
obtain $\xi (r)$. In practice we start with the initial
power variance $\Delta(k)$ to obtain a table of [$k_E$, $\Deltae(k_E)$]. 
An interpolation is then used to carry out the integration
in equation (4). 

The solid curves in Figure 1 show the model predictions for
$\xi(r)$ in the SCDM models
with $\sigma_8=1.24$ and 0.62, and in the FLAT and OPEN models
with $\sigma_8=1$.
These predictions are compared to the results of N-body
simulations shown by the symbols. For SCDM model with
$\sigma_8=0.62$, two different
simulation box sizes are used to show the effects of both finite 
box size and finite simulation resolution on small scales. 
It is clear that the fitting formula works 
well for all cases. We have also obtained results
for other models (i.e. FLAT models with $\Omega_0=0.3$ and 0.1,
OPEN models with $\Omega_0=0.3$ and 0.1, all having $\sigma_8=1$)
and found a similar good agreement between model predictions and simulation
results.

   
\subsection {The Mean Pairwise Peculiar Velocities}

   From the pair conservation equation (Peebles 1980, \S 71),
the ensemble (pair weighted) average of the pairwise peculiar
velocity $\langle v_{12}(r)\rangle 
\equiv \langle [\bv(\bx)-\bv(\bx+\br)]\cdot
{\hat {\br}}\rangle$ can be written as
$$
{\langle v_{12}(r)\rangle \over Hr}=-{1\over 3} {1\over [1+\xi(y,a)]}
{\partial \bxi (y,a) \over \partial \ln a},
\eqno(10a)
$$
where $r$ is the proper, and $y$ the comoving, separation  
between the pairs;
$$
H=H_0\left[\lambda_0(1-a^{-2})+\Omega_0(a^{-3}-a^{-2})+a^{-2}
\right]^{1/2}\eqno(10b)
$$
is the value of Hubble's constant for an expansion factor $a$;
$$
\bxi(y,a)\equiv {3\over y^3}\int_0^y y ^2dy  \xi(y ,a)
         =\int_0^\infty{dk\over k}\Deltae^2(k,a)W(ky).
\eqno(10c)
$$
Thus, in order to obtain $\langle v_{12}(r)\rangle$, we need to work out 
$\partial \Deltae(k,a)/\partial a$. Using the formulae presented 
in Subsection 2.2, it is straightforward to calculate the derivative
$\partial\Deltae/\partial a$. In Appendix A, an explicit    
expression is given for this derivative. 

 Figure 2 shows the comparison between the model prediction
of $\langle v_{12}(r)\rangle$ and the simulation result. 
The agreement between the two is remarkably good for all cases
except for B-SCDM0.62 where simulation result 
is significantly smaller than model prediction
on $r\la \mpch$ (the reason for this will be discussed below).
This agreement shows again that the model of $\xi (r)$ presented in 
\S 2.2 is also valid for describing the time evolution of 
$\xi (r)$. 
Jain (1996) used the time evolution of the two-point correlation
function derived directly from N-body simulations to solve for
$\langle v_{12} (r)\rangle$ from the pair-conservation equation.
The validity of our analytical model shows that 
$\langle v_{12} (r)\rangle$ can easily be calculated from the
initial density spectrum via a single integration.
The discrepancy between model prediction and simulation 
result for the case of B-SCDM0.62 is apparently due to the fact
that this simulation is an intermediate output of SCDM1.24
and has evolved only by a factor of 4.5 in the linear density
growth factor since the initial simulation time. The numerical
artifact of the initial density field generated by the Zel'dovich
approximation should exist to some extent when the system is not
sufficiently evolved.  However, this artifact is expected
to become smaller as the simulation evolves, as
demonstrated by Baugh et al. (1995).  It is therefore gratifying to
see that the agreement between model prediction and simulation result
is indeed better for SCDM1.24 than for B-SCDM0.62.

\subsection {Cosmic Energy Equation} 

The (density weighted) mean square peculiar velocity
of mass particles $\langle v_1^2\rangle $ is related to the two-point 
correlation function by the cosmic energy equation: 
$$
{d\over da} a^2\langle v_1^2\rangle =4\pi G{\bar \rho} a^3
{\partial I_2(a)\over \partial \ln a}, 
\eqno(11)
$$
where ${\bar \rho}$ is the mean density of the universe, and 
$$
I_2(a)\equiv \int _0^\infty ydy\xi(y,a)=\int_0^\infty {dk\over k}
{\Deltae^2(k,a)\over k^2}.
\eqno(12)
$$
Integrating equation (11) once, we have
$$
\langle v_1^2\rangle ={3\over 2}\Omega (a) H^2(a)a^2I_2(a)
\left\lbrack 1-{1\over a I_2(a)}\int_0^a I_2(a )da \right\rbrack.
\eqno(13)
$$
In the linear case, $I_2(a)\propto D^2(a)$ and
$$
\langle v_1^2\rangle ={3\over 2}\Omega (a) H^2(a) a^2 I_2(a)
\left\lbrack 1-{1\over aD^2(a)}\int_0^a  D^2(a )da
\right\rbrack.
\eqno(14)
$$
We have found that equation (14) is a good approximation 
(to an error of $<10\%$) to equation (13) for all realistic 
cases. Thus, for a given cosmogonic model, we can easily obtain
$\langle v_1^2\rangle $.

In Table 2 we compare the values of $\langle v_1^2\rangle$
calculated from equation (14) to those obtained from N-body
simulations. As one can see from equation (12), the contribution
to $\langle v_1^2\rangle$ from the power at small wavenumbers is
not negligible, and the values of $\langle v_1^2\rangle$
derived from the simulations may be sensitive to the loss of
power at $k<2\pi/L$. To see such an effect,
we exclude from our analytic calculations all modes with
$k<2\pi/L$. This is done by setting the lower limit of the integration
on the right hand side of equation (12) to be $2\pi/L$.
The results are given in the left column under the column head
`$\langle v_1^2\rangle^{1/2}$ (model)'. For comparison, the values
in brackets are obtained by assuming $L=\infty$.  
As one can see, the effect of finite box size is indeed significant
when the simulation box size is small. This effect is 
more important for the FLAT and OPEN models than for the SCDM model,
because they have more power on large scales. Table 2 shows that the
agreement between our model predictions and simulation results 
is reasonably good. 

\subsection {Pairwise Peculiar Velocity Dispersion}

The relative velocity dispersion of particle pairs
of separation $r$ is defined as 
$\langle [\bv (\bx)-\bv (\bx+\br)]^2\rangle ^{1/2}$.
In Fig.3 we show (by symbols) the dispersion of the 
pairwise peculiar velocities projected along the separations
of particle pairs
($\langle v_{12}^2(r)\rangle^{1/2}$) in the N-body simulations. 
The main features of $\langle v_{12}^2(r)\rangle^{1/2}$
are (a) monotonic rise at small $r$; (b) saturation
at large $r$; (c) a maximum at medium $r$. 
In Section 2.5.1 we will discuss how these features 
can be explained by some physical arguments. Based on these
arguments, we give (in Section 2.5.2) an empirical fitting
formula for $\langle v_{12}^2(r)\rangle^{1/2}$, so that  
it can be calculated easily for any given power spectrum.
Readers who are mainly interested in using the model
to calculate $\langle v_{12}^2(r)\rangle^{1/2}$ can skip
Section 2.5.1 and go directly to Section 2.5.2.
    
\subsubsection {Behaviour of the pairwise peculiar velocity dispersion}

At large separations where the correlated motion of particle pairs
is negligible, the pairwise peculiar velocity dispersion is
$$
\langle v_{12}^2(r\to\infty)\rangle= {2\over 3}\langle v_1^2\rangle.
\eqno(15)
$$
Thus, the asymptotic value of $\langle v_{12}^2(r)\rangle$
is fixed at a constant for large pair separations. 

For very small separations, the main contribution to the pairwise  
velocity dispersion comes from particle pairs in virialized 
dark matter haloes (e.g. Marzke et al. 1994; 
Sheth 1996; Sheth \& Jain 1996). Assuming 
that dark haloes are spherically symmetric and that  
the velocities of particles in them are isotropic,
we can write
$$\langle v_{12}^2(r)\rangle = {1\over 3}
\langle S^2(\vert\bx\vert)+S^2(\vert\bx+\br\vert)\rangle,
\eqno(16)
$$
where $S(r)$ is the 3D velocity dispersion 
of particles at a radius $r$ from a halo, and 
$\langle\cdot\cdot\cdot\rangle$ denotes the pair-weighted average over
all haloes. Thus, in order to obtain $\langle v_{12}^2(r)\rangle$  
defined in equation (16), we need the density profile,
as well as the mass function, of dark haloes.

The results of N-body simulations (Hernquist 1990;
Navarro, Frenk \& White 1996, hereafter NFW; 
Torman et al. 1996) suggest to write the
density and velocity-dispersion profiles as 
$$n(r)=n_0F(r/r_s), \eqno(17)$$
$$S^2(r)=S_0^2(r_s)G(r/r_s), \eqno(18)$$
where $r_s$ is a scale radius. 
For simplicity, we assume that the velocity dispersion
$S(r)$ at a radius $r$ is $\sqrt{3/2}$ times the circular 
velocity at that radius,
$S^2 (r)= (3/2)G M(r)/r$, where $M(r)$ is the 
mass interior to $r$. Under such assumption $G(x)$ is determined by 
$F(x)$,
$$
G(x)={1\over x}\int_0^x F(x ) x ^2 dx .
\eqno(19)
$$
The total number of pairs of separation $r$
in a halo can then be written as 
$$
{\cal P}(r)dr=8\pi^2n_0^2 r_s^3 r^2dr\int_0^\infty x_1^2dx_1F(x_1)
\int_{-1}^1 F(x_2)d\beta,
\eqno(20)
$$
where
$$x_2^2=x_1^2+(r/r_s)^2-2x_1(r/r_s)\beta.$$
Similarly, the $S^2$-weighted number of pairs in a halo is  
$$
{\cal P}_S(r)dr=8\pi^2n_0^2 r_s^3 S_0^2(r_s) r^2dr
\int_0^\infty x_1^2dx_1 F(x_1)
\int_{-1}^1 F(x_2) [G(x_1)+G(x_2)] d\beta .
\eqno(21)
$$
In carrying out the integrations in equations (20) and (21),
we take $F(r/r_s)=0$ when $r$ is larger than the virial
radius of the halo, $r_v$ [which is defined as the radius
of the mass shell that encloses the halo mass
and settles at about half of its maximum expansion radius,
see e.g. Lahav et al. (1991) for more detailed discussion].      

We use the Press-Schechter formalism (Press \& Schechter 1974) 
to calculate the mass function of dark haloes. In this formalism, 
the comoving number density of dark haloes with mass in the range 
$M\to M+dM$ is
$$
N(M)dM =
-\left({2\over \pi}\right)^{1/2} {{\bar \rho}\over M}
{\deltac \over \sigma (R)}
{d\ln \sigma (R) \over d\ln M}
{\rm exp} \left\lbrack -{\deltac^2 
\over 2 \sigma ^2(R) }\right\rbrack 
{ d M \over M},
\eqno(22)
$$
where $M$ is the mass of the halo. $M$ is related to the initial 
comoving radius $R$ of the region from which the halo formed 
(measured in current units) by 
$M={4\pi \over 3} {\bar \rho}R^3$, $\delta_c$ 
is the threshold overdensity for collapsing. 
The predictions of equation (22) have been tested extensively by
N-body simulations for various cosmogonies 
(e.g. Lacey \& Coles 1994; Mo, Jing \& White 1996).  

With the above discussion, we can now write the pairwise 
velocity dispersion defined in equation (16) as:
$$
\langle v_{12}^2(r)\rangle={1\over 3}
{\int _0^\infty N(M){\cal P}_{S}(r)dM\over  
\int _0^\infty N(M){\cal P}(r)dM}.
\eqno(23)  
$$
To complete the description, we need a model for $\delta_c$ and
the relation between the mass of a halo $M$ and its virial radius $r_v$.
For the values of $\delta_c$ in various cosmologies, 
we take the result summarized in Kochanek (1995; 
see also Bartelmann et al. 1993). 
For the relation between $r_v$ and $M$, we use the formulae
given in Lahav et al. (1991). 
Thus, when $r_s$ and the 
density profile $F(r/r_s)$ are given, we can easily obtain  
$\langle v_{12}^2(r)\rangle$ defined in equation (23).

In this paper we will use the model of NFW for the density profile
of dark haloes:
$$
F(x)={1\over x(1+x)^2} ;
\eqno(24)
$$
$$
\delta_0\equiv {mn_0\over \rho_{\rm crit}} 
= {200\over 3}\left\lbrack \ln(1+c)+(1+c)^{-1}-1\right\rbrack^{-1}
c^3,
\eqno(25)
$$
where $\rho_{\rm crit}$ is the current value of the critical density,
$m$ is the mass of a particle and $c$ is the concentration factor,
$$
c\equiv {r_{200}\over r_s},
\eqno(26)
$$
with $r_{200}$ being the radius within which the average mass overdensity
of the halo is 200. It should be pointed out that $r_{200}$ 
is in general different from the virial radius $r_v$, but
for a given density profile the relation between $r_v$
and $r_{200}$ is fixed.
The exact value of $c$ for a halo depends on its formation history
and cosmology. The simulation result of NFW shows that $c\sim 5$-10
for typical haloes [i.e. those with Lagrangian radius 
$R\sim R_*$ where $R_*$  
is defined by $\sigma(R_*)=1.69$] in the
standard CDM model. NFW also proposed a simple model
for $c$ based on halo formation time. The formation redshift
$z_f$ of a halo identified at redshift $z<z_f$ with mass 
$M$ is defined as the redshift by which 
half of its mass is in progenitors with mass exceeding $fM$, where
$f<1$ is a constant. According to the formula given by Lacey \& Cole
(1993) based on the Press-Schechter formalism, the halo formation time
is then defined implicitly by
$$
{\rm erfc}\left\lbrack {(\delta_{z_f}-\delta_z)
\over \sqrt {2[\sigma^2(fM)-\sigma^2(M)]}}\right\rbrack={1\over 2},
\eqno(27)
$$
where $\delta_z$ is the threshold overdensity for collapsing at
redshift $z$ [e.g. $\delta_z=(1+z)\delta_c$ in an Einstein-de Sitter
universe], $\sigma (M)$ is the rms of the linear power spectrum 
at redshift $z=0$ in top-hat windows enclosing mass $M$ 
(see equation 2). NFW have used $M_{200}$ (the mass enclosed
in $r_{200}$) instead of $M$ in equation (27). We will follow their
convention.
NFW suggested that the characteristic overdensity of a halo identified
at redshift $z$ with mass $M$ is related to its formation redshift
$z_f$ by
$$
\delta_0(M,f,z)=C(f)\Omega_0 \left\lbrack 1+z_f(M,f,z)\right\rbrack ^3,
\eqno(28)
$$
where the normalization $C(f)$ depends on $f$. We will take $f=0.01$
as suggested by the N-body results of NFW. In this case $C(f)\approx
2\times 10^3$ (NFW). 
Thus for a halo of given mass, one can obtain the
concentration factor $c$ from equations (25)-(28). 
In practice, we first solve $z_f$ from equation (27) and insert 
the value of $z_f$ into equation (28) to get $\delta_0$,
we then use this value of $\delta_0$ in equation (25) to solve for $c$. 
As shown by NFW, the value of $c$ does not depend sensitively 
on the choice of $f$, as long as $f\ll 0.1$.
It is necessary to point out that the arguments given in 
NFW are originally only for Einstein--de Sitter universe. 
Fortunately, these arguments are equally valid for
low-$\Omega$ universes, as is shown recently by Navarro et al. 
(1996, in preparation). 

The dotted lines in Fig.3 show $\vdisper$ given by equation (23)
as a function of $r$. The pairwise velocity dispersion 
$\vdisper$ increases with $r$, because for large $r$ the fraction 
of pairs from big haloes is larger. To examine the sensitivity of 
our results to the choice of the model for the concentration 
factor $c$, we also made calculations for the SCDM models by assuming 
$c=5$ and 10. The results are shown in Figs. 3a and 3b as 
the dashed curves (how these curves are obtained will become
clear in the following). A larger value of $c$ gives higher  
amplitudes for $\vdisper$, because an increase of $c$ enhances
the velocity dispersion of a halo in the central region.  

It is clear that model (23) will break down for large $r$
where the number of pairs within virialized haloes becomes smaller
than other kinds of pairs (e.g. cross pairs between halo
particles and field particles, and pairs among field particles).  
At some separation, the value of $\langle v_{12}^2\rangle$ 
should decrease with $r$ and approach the asymptotic
value given by equation (15). Although the function
$\langle v_{12}^2(r)\rangle$ in the medium range 
of $r$ is complicated, (because it depends not only on the
amplitude but also on the shape of the three-point correlation 
functions, see below), it may still be possible to find some 
physically motivated models to understand the features
observed in the simulations, given the asymptotic bahaviour
of $\langle v_{12}^2(r)\rangle$ on both small and large 
separations. 

We start with equation (72.1) in Peebles (1980):
$$
{\partial \over \partial t}
\left(1+\xi\right)v^{\alpha} 
+{{\dot a}\over a}\left(1+\xi\right)v^{\alpha}
+{1\over a}{\partial \over \partial x^{\beta}}
\left(1+\xi\right)\langle {v_{12}}^{\alpha} {v_{12}}^{\beta}\rangle
$$
$$
+{2Gm\over a^2}{x^{\alpha}\over x^3}\left(1+\xi\right)
+2G{\overline \rho}a{x^{\alpha}\over x^3}\int_0^x d^3x \xi
+2G{\overline \rho}a\int d^3x_3\zeta(1,2,3)
{{x_{13}}^{\alpha}\over x_{13}^3}=0 ,
\eqno(29)
$$
where $v^{\alpha}=\langle {v_{12}}^{\alpha}\rangle$ is the 
$\alpha$-component of the mean pairwise peculiar velocity 
(see equation 10a), $\zeta (1,2,3)$ is the three point 
correlation function, $m$ is the mass of a particle.   
The fourth term describes the mutual attraction of the particle 
pair and can be neglected when we consider dark matter particles.
We write the velocity dispersion as 
$$
\langle {v_{12}}^{\alpha} {v_{12}}^{\beta}\rangle 
=\left[{2\over 3} \langle v_1^2\rangle +\Sigma\right]
\delta_{\alpha\beta}+\left[\Pi -\Sigma \right] 
{x^{\alpha} x^{\beta} \over x^2},
\eqno(30)
$$
where $\langle v_1^2\rangle$ is the mean square peculiar 
velocity of particles (equation 11), $\Pi$ and $\Sigma$ represent
the effects of the correlated motions on the components of the
dispersion parallel and perpendicular to the separation.
Thus the (1D) velocity dispersion defined in the beginning of this 
subsection is
$\vdisper={2\over 3}\vone +\Pi$. Using the boundary condition
that $\xi(r\to \infty)= 0$ and $\Pi (r\to \infty)= 0$
(see Peebles 1980, \S 72), we can obtain a formal 
solution for $\vdisper$ from equation (29):
$$
\left[1+\xi(r)\right]\vdisper
={2\over 3}\vone
+3\Omega (Hr)^2I
-(Hr)^2\left[{\partial ^2 I\over \partial (\ln a)^2}
+{\partial I\over \partial \ln a}{\partial \ln H\over\partial \ln a}
+2{\partial I\over \partial \ln a}\right]$$
$$+2\int_r^\infty {dr \over r }\left[1+\xi(r )\right](\Pi-\Sigma)
+{3\Omega H^2\over 4\pi} \int_r^\infty {dr \over r }
\int d^3q{{\bf r }\cdot {\bf q} \over q^3} 
\zeta\left(r ,q,\vert {\bf r }-{\bf q}\vert\right),
\eqno(31)
$$
where
$$
I\equiv \int_0^\infty {dk\over k}\Delta_E^2(k,a){\sin kx\over (kx)^3}.
\eqno(32)
$$
Note that $r$ in equation (31) is the physical radius.  
Since models for $\xi$, $\vone$ and $\Delta_E$ have already been 
constructed,  we need only to specify $\Pi-\Sigma$ and the 
integration of $\zeta$ in the last term of equation (31) to 
complete our model for $\vdisper$. 
  
For our demonstration, we will assume that the three point 
correlation function $\zeta$ on small scales obeys the hierarchical form:
$$
\zeta(r_1,r_2,r_3)=Q\left[\xi(r_1)\xi(r_2)+\xi(r_2)\xi(r_3)
+\xi(r_3)\xi(r_1)\right],
\eqno(33)
$$
where $Q$ is a constant. It is clear from equation (31) that the
pairwise peculiar velocity dispersion on small scales is dominated by the
term of the three point correlation. In this case, we can write
$$
\vdisper ={3\Omega H^2Q\over4\pi [1+\xi(r)]}\int_r^\infty
{dr \over r }\int d^3q {{\bf r }\cdot {\bf q} \over q^3} 
\left[\xi(r )\xi(q)+\xi(q)\xi(\vert {\bf r }-{\bf q}\vert)
+\xi(\vert {\bf r }-{\bf q}\vert)\xi(r )\right].
\eqno(34)
$$
Since $\xi$ is known for a given model (see \S 2.2),
the value of $Q$ can then be determined by specifying 
$\vdisper$ at some small fiducial radius $r_Q$. 
On using our model for $\vdisper$ on small scales, the last term of
equation (31) can therefore be fixed, as long as the hierarchical 
form (33) holds. On small scale, the hierarchical form is 
a reasonable approximation, however on large scales, $Q$ is not a constant
but depends on the size and shape of the triangle specified
by ($r_1,r_2,r_3$) (see e.g. Matsubara \& Suto 1994; Jing \& B\"orner 1996). 
Unfortunately, a theoretical model is not yet available 
to describe $Q$ as a function of ($r_1,r_2,r_3$)
{\footnote {Note that the relevant quantity 
in our problem is the integration of $\zeta$ in the last term 
of equation (31), which is only a one-dimensional function of $r$   
and should therefore be much easier to model than $\zeta$ itself.}.     
In what follows, we assume equation (33) to hold on all
scales. We will show that the error in $\vdisper$
caused by such an assumption can be corrected by a simple
model.    

To model $\Pi-\Sigma$, we define a quantity which describes 
the anisotropy of the relative peculiar velocity dispersion: 
$$
{\cal A} (r)\equiv 1-{{{2\over 3}\vone +\Sigma (r)\over \vdisper}}.
\eqno(35a)
$$
It then follows that
$$
\Pi-\Sigma={\cal A}(r)\vdisper .
\eqno(35b)
$$
On small scales where the velocity dispersion is isotropic, we 
have ${\cal A}=0$. On large scales where linear theory applies, 
we can write (see Peebles 1993, \S21, note also the slight 
difference in the definitions of $\Pi$ and $\Sigma$):   
$$
\Pi (r)=-{2\over 3}\left({{\dot D}\over D}\right)^2
\left[{J_5(r)\over r^3}+K_2(r)\right];
\eqno(36a)
$$ 
$$
\Sigma (r)= -\left({{\dot D}\over D}\right)^2
\left[{J_3(r)\over r}-{J_5(r)\over 3r^3}+{2\over 3} K_2(r)\right],
\eqno(36b)
$$
where
$$
J_3(r)=\int_0^r r^2\xi(r)dr=r^3\int_0^\infty {dk\over k}
{\Delta^2(k)\over (kr)^3}\left(\sin kr-kr\cos kr\right);
$$
$$
J_5(r)=\int_0^r r^2\xi(r)dr=r^5\int_0^\infty {dk\over k}
{\Delta^2(k)\over (kr)^5}\left\lbrace kr\left[6-(kr)^2\right]\cos kr
+3\left[(kr)^2-2\right]\sin kr\right\rbrace ;
$$
$$
K_2(r)=\int_r^\infty  r\xi(r)dr=r^2\int_0^\infty {dk\over k}
{\Delta^2(k)\over (kr)^2}\cos kr.
$$
In the above equations, $\Delta ^2(k)$ is the dimensionless power 
spectrum in the linear regime. We will, however, insert the
nonlinear power spectrum so that some nonlinear effects
can be taken into account. 
 Figure 5 compares ${\cal A}$ predicted by equations (35) and (36)
(solid curves) to that derived from the simulations.
The overall anisotropy in the velocity dispersion is
quite small, and on separations $r\gsim 2\mpch$ it is reasonably 
well described by our model. Since the $(\Pi-\Sigma)$ term is 
expected to be important   
only on large scales, we will use equation (36) to estimate
($\Pi-\Sigma$). 

 The dot-dashed curves in Fig.3 show the pairwise peculiar velocity
dispersion $\vdisperr$ given by equation (31) with the assumptions
discussed above. We have taken $r_Q=0.1\mpch$. As one
can see from the figure, the result is not sensitive 
to the exact value of $r_Q$, as long as $r_Q$ is small.
Our model successfully gives a maximum in $\vdisperr$,
as observed in the N-body simulations. The maximum
occurs at a separation characteristic of the size
of typical clusters, about 1-$2\mpch$. This happens because most
particles forming pairs of larger separations are in 
low-density environments where the velocity dispersion is low.
The model overestimates the pairwise peculiar
velocity dispersion on large scales when compared to the results
derived from the N-body simulations (shown by the symbols). 
This overestimation comes mainly from our assumption that 
$Q$ is a constant. Indeed, the value of $Q$ may decrease   
substantially on large scales, as shown by N-body simulations
(e.g. Jing \& B\"orner 1996). The contribution to 
$\vdisper$ from the three-point correlation is therefore 
overestimated on large scales in our model. As discussed above, 
a reliable model for $\zeta$ over different scales is still   
lacking at present time, and we have to invoke some
simple arguments to correct the errors caused by the assumption
of equation (33). 

 As we have discussed above, for small separations $r$, 
most pairs are in virialized haloes and equation (33) should be a 
reasonable description. The assumption of equation (33) fails
for large $r$, because pairs outside dark haloes become
important. As the simplest choice,
the fraction of particles in dark haloes with
radius $r_{200}\ge \alpha_h r$ ($\alpha_h$ being a constant), 
$f_h(\alpha_h r)$, may be a good guess of the function that 
controls the change of $\langle v_{12}^2(r)\rangle$ from 
its small separation value to its large separation value. 
According to the Press-Schechter formalism (see equation [22]),
$f_h(\alpha_h r)$ can be written in terms of the initial power
spectrum as:
$$
f_h(\alpha_h r)=\erfc\left\lbrack {\delta_c\over \sqrt{2}
\sigma(R_h)}
\right\rbrack,
\eqno(37)
$$
where $R_h$ is the linear radius of a halo with $r_{200}=\alpha_h r$. 
Thus when $r$ is very small, all particles are in haloes so that 
$f_h=1$, while $f_h=0$ when $r\to  \infty$. This behaviour of
$f_h$ suggests the following empirical formula for   
$\langle v_{12}^2(r)\rangle$:
$$
\langle v_{12}^2(r)\rangle=
[f_h(\alpha_h r)]^{\tau}\langle v_{12}^2(r)\rangle_s
+\left\lbrace 1-[f_h(\alpha_h r)]^{\tau}\right\rbrace\times {2\over 3}
\langle v_1^2\rangle ,
\eqno(38)
$$
where we have changed our notation so that 
$\langle v_{12}^2(r)\rangle_s$ denotes the $\langle v_{12}^2(r)\rangle$
defined by equation (31). The values of $\alpha_h$ and $\tau$
are to be calibrated by results from N-body simulations.

 The thick solid curve in Figure 3a is calculated according to
(38) with the values
$$
\alpha_h=0.2; \,\,\,\, \tau=1 .
\eqno(39)
$$
It is clear that the model fits the simulation results 
for the SCDM model with $\sigma_8=0.62$ reasonably well. 
The predictions of $\vdisper$ for other cosmogonical models
are shown by the thick curves in the other panels of Fig.3,
using the same values of $\alpha_h$ and $\tau$ as 
given in equation (39). 
The figure shows that our model also works for these cosmogonies. 
For comparison, the two dashed curves in Figs. 3a and 3b
show the same results obtained by assuming the concentration
factor $c=5$ (lower curve) and 10.

The good agreement between the model predictions and the N-body
results shows that the main features in $\langle v_{12}^2(r)\rangle$
can indeed be explained by our simple physical 
considerations. 
%
%
%
%

\subsubsection {Fitting formula for the pairwise peculiar velocity dispersion}

The model for $\vdisper$ given in Section 2.5.1, although 
having physical motivations, is not easy to
implement. In this section we provide a much simpler 
fitting formula for $\vdisper ^{1/2}$.

 Based on the N-body results and the discussion in Section 2.5.1 we
make the following ansatz for $\vdisper$:
$$
\vdisper ^{1/2}=\Omega^{0.5}Hr_c\phi(r/r_c),
\eqno(40a)
$$
where $\phi(x)$ is a universal function and $r_c$ is a nonlinear
scale. We choose $r_c$ to be the virial
radius of $M^*$ haloes, $r_v^*$. For a given power spectrum,
the linear radius of $M^*$ haloes, $r_0^*$ , is given 
by $\sigma (r_0^*)=1$. The relation between $r_v^*$ and 
$r_0^*$ in different cosmological models can be obtained by using the 
formulae given in Appendix B. We approximate the functional form
of $\phi (x)$ by:
$$
\phi(x)={\phi_\infty (1+Bx^{-\beta})+Ax^{-\alpha}
\over 1+V[g(a)]^{0.35}
[\Omega(a)]^{0.2}x^{-(\alpha+\epsilon)}} ,
\eqno(40b)
$$
where $\phi_\infty 
=\sqrt{{2\over3}}\langle v_1^2\rangle^{1/2}
/(Hr_c\Omega^{0.5})$; $A$, $B$, $V$, 
$\alpha>\beta>0$, and $\epsilon>0$ are constant.
It follows that for $r\to \infty$, $\vdisper$ is forced to have the 
asymptotic value given by equation (15). 
For $r\to 0$, $\vdisper ^{1/2}\propto x^{\epsilon}$
so that it increases with $r$ as a power law. 
The dependence on the linear growth factor $g(a)$ is included 
to take into account the fact that for a given power spectrum,
haloes with the same linear radius may have different density
profiles (or concentrations) in different cosmological models 
(see Section 2.5.1). The power of $g(a)$ can be fixed by 
comparing the amplitudes of $\vdisper ^{1/2}$ at small $r$
for models (having the same initial power spectrum) with 
or without a cosmological constant. We found that a value
of about 0.35 gives a reasonably good fit to our N-body results.
The dependence on $\Omega$ is included because for a given
virial radius of a halo the ratio $S/\Omega^{0.5}$
(where $S$ is the velocity dispersion of the halo) 
is larger in a lower-density universe.

The solid curves in Fig.4 show the predictions of our
fitting formula with
$$
A=58.67;\,\,\, B=-0.3770;\,\,\, V=4.434;
\eqno(40c)
$$
$$
\alpha=2.25;\,\,\, \beta=1.90;\,\,\, \epsilon=0.15 .
\eqno(40d)
$$
It is clear that the formula gives a reasonably good fit
to all cosmogonical models examined in the present paper.
At the moment, it is not clear whether the small discrepancy
(typically $\la$ 20 percent) between the model predictions and
simulation results in some cases is due to the inaccuracy of 
our model, or due to the lower resolution of our simulations,
or due to both of them.
As one can see from Fig.4a, a lower resolution indeed
reduces the pairwise peculiar velocity dispersion. 
Obviously, better simulations are needed to give a more accurate 
calibration of our formula.  
Given the uncertainties in our present simulations, 
we do not intend to obtain the best fit values
for the parameters in equation (40b).
   
\subsection 
{The Hierarchical Amplitude of the Three-Point Correlation Function}

 Given the model for $\vdisper$, we can use equation (34) to 
calculate the hierarchical amplitude $Q$ on small scales. 
In Figure 6 the solid curves show $Q$ as a function of $r$ predicted 
by our fitting formula (40). The amplitude $Q$ decreases only slowly with
$r$ for $r\sim 0.1 \mpch$, showing that equation (33) is valid
on small scales. The symbols in the figure show the values 
of $Q$ estimated from N-body simulations by averaging over all
triplets with the smallest side equal to $r$ and with the second
smallest side less than $4r$. Thus this quantity
is {\it not} exactly the same as the one given by the model. However,
since for small $r$ the value of $Q$ does not depend strongly
on the shape of the triplets (see Jing 1996, in preparation),
the comparison between the model and the simulation results
may still be meaningful. The figure shows that our model prediction is 
generally in agreement with the simulation results. 
The values of $Q$ at small separations ($r\sim 1\mpch$) are typically
1-2, without a strong dependence on cosmogonies. These values 
of $Q$ are only slightly  larger than those obtained for galaxies
($Q\sim 1$ for optical galaxies, see e.g. Groth \& Peebles 1977;
Jing, Mo \& B\"orner 1991). 

 If the cosmic virial theorem is used under the assumptions 
that the two-point correlation function has a power-law form,  
$\xi(r)=A_0r^{-\gamma}$,  and that
the three-point correlation function $\zeta$ has the hierarchical
form given by equation (33), then the pairwise velocity dispersion 
on small scales can be written as: 
$$
\langle v_{12}^2(r)\rangle=
{2\pi G{\bar \rho}Q A_0r^{2-\gamma} J(\gamma)\over 
(\gamma-1)(2-\gamma)(4-\gamma)}
\eqno(41)
$$
(see Peebles 1980, Section 75).
Here
$$J(\gamma)=\int_0^\infty dyy^{-2}(1+y^{-\gamma}){\cal M}$$
with
$$
{\cal M}=(1+y)^{2-\gamma}\left[1-(2-\gamma)y+y^2\right]
-\vert 1-y\vert ^{2-\gamma}\left[1+(2-\gamma)y+y^2\right].
$$
Thus, by using our model for $\langle v_{12}^2(r)\rangle$
and by fitting $\xi(r)$ to a power law on small scales to get
$A_0$, we can also get the value of $Q$ from equation (41).  
The validity of equation (41) was checked by Suto (1993) using
N-body simulations.
There is, however, an uncertainty here. As one can see
from Fig.1, $\xi(r)$ is not a good power-law even on small scales,
the value of $\gamma$ determined from $\xi$ 
depends therefore on the range of $r$ used in the fitting.
Indeed, for $r\sim 1\mpch$, the local value of $\gamma$
can sometimes be larger than 2 and equation (41) is ill-defined.
Since it is unclear which range of $r$ is most relevant
for $\vdisper$ on small scales, it does not seem to be 
a good idea to use $\gamma$ obtained from $\xi$ in equation (41)
to obtain $Q$. Fortunately, the value of $\gamma$ in equation (41)
can also be determined from the dependence of
$\vdisper$ on $r$. On small scales, $\vdisper$ increases
with $r$ approximately as a power law (Fig.4). Thus the value of
$\gamma$ given by fitting $\vdisper$ to a power law
is always smaller than 2, and equation (41) is well defined.
The value of the correlation amplitude, $A_0$, can still
be determined reasonably well by fitting $\xi(r)$ on small scales 
to a power law. As one can see from equation (41), 
the error in $Q$ induced by $A_0$ is only proportional 
to that in $A_0$.   
In Table 2, we show the values of $Q$ predicted 
by equations (40) and (41) for $\xi(r)$ and 
$\langle v_{12}^2 (r)\rangle$ in the
range $r=0.1-0.5\mpch$, along with the values of $\gamma$ 
obtained by fitting $\vdisper$ to a power law. The values of 
$Q$ obtained in this way are similar to those shown in Fig.6.

%
%

\section {THE EFFECT OF REMOVING MASSIVE HALOES ON 
    $\langle v_{12}^2(r)\rangle$} 

 Davis \& Peebles (1983) have calculated the pairwise velocity 
dispersion of galaxies for the CfA1 redshift survey (Huchra et al.
1983). Their result, $\langle v_{12}^2(r)\rangle ^{1/2}
=340\pm 40\kms$, had been used for about 10 years to judge 
models of structure formation, and was a primary argument against 
the CDM model with $\Omega_0=1$ and with $\sigma_8\sim 1$. 
Based on the same data set, 
Mo, Jing \& B\"orner (1993, see also Zurek et al. 1994;
Somerville, Davis \& Primack 1996)
found that this statistic is sensitive to the presence (or absence) 
of galaxy clusters in a sample, and the constraint given 
by the observational results is uncertain.
Recently, similar analyses have been performed on new, larger
redshift surveys (Fisher et al. 1994; Guzzo et al. 1995; 
Marzke et al. 1995), but the problem remains: there is a large
variation in $\langle v_{12}^2(r)\rangle$ between different 
surveys, and even for the same survey analysed in different ways. From
our model for $\langle v_{12}^2(r)\rangle$ (see Section 2.5.1), 
we see clearly why the observed small scale pairwise velocity dispersion
is sensitive to the presence (or absence) of galaxy clusters.   
Indeed, from equations (20)-(23) we can write 
$$
\langle v_{12}^2(r)\rangle\propto
{\int _0^\infty N(M) M^{5/3} dM\over  
\int _0^\infty N(M)MdM},
\eqno(42)  
$$
where we have assumed that $M\propto r_s^3$ and 
$S_0^2\propto M^{2/3}$. These assumptions are approximately
correct for haloes with density profiles close to isothermal.  
Using equation (22) we see
that while the number of pairs with small separations  
is dominated by dark haloes with $M\sim M_*$ (where $M_*$ is the mass
at which the rms mass fluctuation $\sigma= 1$),
the pairwise peculiar velocity dispersion on small scale
can be significantly affected by massive ones with $M>M_*$. 
The number density
of such massive haloes is small and a large sample 
is therefore needed to have a fair sampling of their 
number density. Marzke et al. (1995) have discussed 
in considerable detail how such sampling can 
affect the statistics on $\vdisper$. In this paper, 
we will not discuss this effect in detail.
Instead, we show that from our model one can derive some
new statistics which may be more robust against the sampling
effect discussed above. 

  Somerville, Primack \& Nolthenius (1996) have suggested that
studying $\vdisper$ as clusters (dark haloes) with 
different internal velocity
dispersions are removed leads to interesting information about
the amount of power on cluster and subcluster scales. 
Since in practice the internal velocity dispersions of dark 
haloes are not easy to measure accurately for a large number
of haloes, we suggest to study $\vdisper$ as a function
of the mean separation of the clusters that are removed 
from the analysis. 

  From equation (22), we can write the total comoving number
density of haloes with mass exceeding $M$ as
$$
N(>M)=-{3\over (2\pi)^{3/2}}\int_R^\infty
{1\over R^3}{\delta_c\over \sigma (R)}
{d\ln \sigma\over d\ln R}{\rm exp}
\left[-{\delta_c^2\over 2\sigma^2}\right] {dR\over R}.
\eqno(43)
$$ 
The mean separation of haloes with mass exceeding $M$
is therefore ${\overline d}(M)\equiv [N(>M)]^{-1/3}$. As one can see from
equation (43), for a given power spectrum, ${\overline d}$
determines $R$. Thus the effect of removing haloes 
with masses exceeding $M$ in the model is to change 
the upper limits of the integrations in equation (23)
from $\infty$ to $M$. This obviously has the effect of 
reducing $\vdisper$ on small scales. In Figure 7 we
show $\langle v_{12}^2(r)\rangle^{1/2}$ at 
$r=0.1\mpch$ as a function of ${\overline d}$ for 
the SCDM models ($\Gamma=0.5$) with $\sigma_8=0.62$ and 1.24, and for 
three other
cosmogonical models with $\Gamma=0.2$ and 
$\sigma_8\Omega_0^{0.6}=0.38$. 
The use of the combination $\sigma_8\Omega_0^{0.6}$
is motivated by the fact that this quantity 
can be determined from observations based on large scale 
velocity fields (see e.g. Dekel 1994, Fisher et al. 1994) and 
on cluster abundance
and clustering (e.g. White, Efstathiou \& Frenk 1992;
Mo, Jing \& White 1996). The specific value $0.38$ is chosen simply 
because the OPEN and FLAT models have this value. 
The figure shows that the value
of $\langle v_{12}^2(r)\rangle^{1/2}$ does not reach its asymptotic
value even for ${\overline d}\sim 50\mpch$ (the typical value
for the mean separation of rich clusters). Thus to have 
the small scale pairwise peculiar velocity dispersion fairly sampled,
one needs a sample that contains many rich clusters. All galaxy redshift
samples used so far to derive $\vdisper$ are not large enough to 
qualify for this purpose, and it is not surprising that the values of
the small scale pairwise peculiar velocity dispersion of galaxies
derived from available galaxy samples are uncertain. 

 However, by inspecting Fig.7 we see that some useful statistics based on the
pairwise velocity dispersion can still be derived from relatively
small samples. For a given sample, we first identify from it
clusters (groups) of different masses. We then remove 
successively the most massive clusters from the sample and
analyze the small scale pairwise peculiar velocities as a function
of the mean separation of clusters (${\overline d}$) 
that have been removed from the sample. Given a small sample,
such a function can be determined reliably only 
for small values of ${\overline d}$,
because the mass function is well sampled only for
relatively poor clusters. As one can see from Fig.7, the
model prediction for
this function is quite different for different cosmogonies.

 Another interesting point to note from Fig.7 is the difference
in the results for the OPEN, FLAT and the 
($\Omega_0=1, \Gamma=0.2, \sigma_8=0.38$) 
model. All these three models have the same shape of the power
spectrum ($\Gamma=0.2$) and the same value of 
$\sigma_8\Omega_0^{0.6}$ ($=0.38$). As we have discussed before, 
it is difficult to determine $\Omega_0$ and $\sigma_8$ separately from
observations on large scale velocity fields or on cluster
abundance. On the other hand, it is relatively easy to 
determine the combination $\sigma_8\Omega_0^{0.6}$ from these
observations. Fig.7 shows, however, that one can hope to 
separate $\Omega_0$ and $\sigma_8$ by measuring the small scale 
peculiar velocity dispersion. The small scale pairwise 
peculiar velocity dispersion
is lower in the model with $\Omega_0=1$ than in the OPEN and FLAT
models with the same $\sigma_8\Omega_0^{0.6}$, because 
clusters in this model formed late and thus are less concentrated.    
For the same reason, the small scale pairwise peculiar velocity
dispersion is lower in the FLAT model than in the OPEN model.
This line of argument is obviously worth further investigation.
  

\section {DISCUSSION}

  The analytical models presented in Section 2 allow
us to see clearly how the low order statistics of the density
and the velocity distributions are determined by
cosmological model and initial power spectrum. As a result 
we can use them to construct some statistical
measures of the density and peculiar velocity fields
to constrain models by observations. In this Section we will 
briefly discuss several possible applications of our results.
Details of these applications will be described elsewhere.

  As a first application our results can be used to construct
models for the reconstruction of cosmogonical parameters from 
measurements in redshift space. As discussed in Fisher et al.
(1994, and references therein), 
in order to use the redeshift-space distortion on the two-point
correlation function to derive cosmological parameters,
it is necessary to model the distribution of the pairwise peculiar
velocity. The distribution function is usually constructed
from the low-order statistics of the pairwise peculiar velocity,
which, by using our model, can all be calculated directly for 
any given cosmogony. Thus a liklihood function can be written
for the pairwise peculiar velocities, which depends only on the
cosmogonical parameters. Given a redshift sample of galaxies,
one can then hope to constrain the cosmogonical parameters
by maximizing the likelihood function. 

 A second application of our models is to use them to examine what
happens to some statistics, if the power spectrum is filtered
in some manner. Such an investigation is important, because
(1) real observations do not sample all parts of the spectrum
equally well, and we need to understand how such sampling affects
our statistics; and (2) we may sometimes want to filter the density
field deliberately in order to get some insight into the underlying density 
field. In Section 3 we have already shown one such example related
to the effect of removing massive clusters on the values of
the pairwise peculiar velocity dispersions on small scales. Here
we give another one.

 The mean square velocity $\vone$ given by the cosmic energy equation 
presented in Subsection 2.4 contains both large scale bulk motion
and small scale random motion. It is sometimes desirable to separate
these two kinds of motions by filtering the mass density
spectrum. From 
equation (13), one can have a `filtered version' of the 
cosmic energy equation:      
$$
\langle v_1^2(x)\rangle ={3\over 2}\Omega H^2a^2
\left\lbrack I_2(x,a)-{1\over a}\int_0^a I_2(x,a)da\right\rbrack,
\eqno(44)
$$
where
$$
I_2(x,a)=\int_0^\infty {dk\over k}
{\Deltae^2(k,a)\over k^2}W(k;x)
\eqno(45)
$$
with $W(k;x)$ being a filter with characteristic (comoving)
radius of $x$. It is clear from equation (45) that if the 
high-frequency modes are filtered out from the 
dimensionless power spectrum $\Delta^2_E$, then the velocity defined by
equation (44) describes the motions induced by 
the low-frequency modes of the density perturbations. 
In linear case, this is just the mean square of the bulk motion 
of dark matter in the filter. On the other hand, if low
frequency modes are filtered out, the velocity defined by
equation (44) describes small scale relative motions. 
The statistics of the filtered velocity field can be derived either from 
velocity data (e.g. those based on Tully-Fisher relations)
or from redshift-space distortions (see Miller, Davis \& White 1996 
for a discussion). Our model will then provide a clear picture
how cosmogonical models are constrained by these measurements.

It must be pointed out once again that in order to make any
rigorous comparison between models and observations of galaxy
distribution we need to understand how galaxy distribution is related
to dark matter distribution. Much work needs to be done in this
respect.  However, for a given cosmogonical model, the density bias
parameter, defined by the ratio between the two-point correlation
function of galaxies and that of mass, and the velocity bias
parameter, defined by {\it e.g.} the ratio between the pairwise
peculiar velocity dispersion of galaxies and that of dark matter, are
actually fixed.  Our model can therefore be used to
understand the effects of biased galaxy formation on the low-order
moments of galaxy distribution with respect to those of dark matter.

Finally, it is necessary to point out that the accuracy of our
analytical models depends on the accuracy of the N-body 
simulations used to calibrate them. 
At present time, the discrepancy between our
analytical formulae and simulation results is typically 20 
percent. It is still unclear whether this small but significant 
discrepancy results from the inaccuracy of our analytical models, 
or from the limited resolution of N-body simulation, or from both
of them. However, the method described in our paper is general. 
With improved N-body simulations, the fitting formulae for the 
evolved power spectrum and for the halo density profiles can 
all be improved. More accurate models for the low-order moments 
of dark matter distribution can then be easily obtained by using 
the procedure outlined in our paper.

\acknowledgments
We thank John Peacock, our referee,
for a very useful report.
We also thank Julio Navarro, Yasushi Suto, David Syer and Simon White 
for helpful discussions. The simulation data was generated when
YPJ held an Alexander-von-Humbodlt research fellowship.
This work was also supported by SFB375.

\appendix
\section{The Explicit Expression for $\partial \Delta_E^2/\partial a$}

  From equations (5)-(9) we see that $\Delta_E^2(k,a)$
depends on the expansion factor $a$ through $g(a)$ and
$x\equiv \Delta^2(k_L)$, and we can write
$$
\left({\partial \Delta_E^2(k,a)\over \partial a}\right)_k
={\partial f\over \partial g}{dg\over da}
+{\partial f\over \partial x}
\left({\partial x\over \partial a}\right)_k .
\eqno(A1)
$$
The linear power spectrum $\Delta^2$ evolves according to the
linear theory so that
$x=[{\cal G}(a)]^2x_0$, where ${\cal G}(a)\equiv [ag(a)/a_0g(a_0)]$ and
$x_0\equiv \Delta^2(k_L,a_0)$ is the linear power spectrum
at a fiducial expansion factor $a_0$ (e.g. at present time).
Thus
$$
\left({\partial x\over \partial a}\right)_k
={d{\cal G}^2\over da}x_0+{\cal G}^2 {dx_0\over dk_L }
\left({\partial k_L\over \partial a}\right)_k,
\eqno(A2)
$$
where $k_L\equiv [1+\Delta_E^2(k,a)]^{-1/3}k$.
Inserting (A2) into (A1) we have:
$$
\left({\partial \Delta_E^2(k,a)\over \partial a}\right)_k
=\left\lbrack {\partial f\over \partial g}{dg\over da}
 +{\partial f\over \partial x}{d{\cal G}^2\over da} x_0
\right\rbrack
\left\lbrack 1+{{\cal G}^2\over 3}\left({k_L\over k}\right)^3
{dx_0\over d\ln k_L}{\partial f\over \partial x}
\right\rbrack ^{-1}.
\eqno(A3)
$$
The derivatives on the right hand side of (A3) are:
$$
{dg\over da}={\partial g\over \partial \Omega}{d\Omega\over da}+
{\partial g\over \partial \lambda}{d\lambda\over da};
\eqno(A4)
$$
$$
{d\Omega\over da}
=-{\Omega_0(1-\Omega_0-\lambda_0+3a^2\lambda_0)
\over \left[a+\Omega_0(1-a)+\lambda_0(a^3-a)\right]^2};
\eqno(A5)
$$
$$
{d\lambda\over da}
={a^2\lambda_0[2a(1-\Omega_0-\lambda_0)+3\Omega_0]
\over \left[a+\Omega_0(1-a)+\lambda_0(a^3-a)\right]^2};
\eqno(A6)
$$
$$
{\partial g\over \partial \Omega}=
{15\Omega^{4/7}/14+{5/2}-69\lambda/28
\over \left\lbrack \Omega^{4/7}
-\lambda +\left(1+{\Omega/ 2}\right)
\left(1+{\lambda/ 70}\right)\right\rbrack ^2}; 
\eqno(A7)
$$
$$
{\partial g\over \partial \lambda}=
{(5\Omega/2)\left\lbrack 1-
\left(1+{\Omega/ 2}\right)/70\right\rbrack
\over \left\lbrack \Omega^{4/7}
-\lambda +\left(1+{\Omega/ 2}\right)
\left(1+{\lambda/ 70}\right)\right\rbrack ^2}; 
\eqno(A8)
$$
$$
{d{\cal G}^2\over da}=
{2{\cal G}^2\over a}\left(1+{d\ln g\over d\ln a}\right);
\eqno(A9)
$$
$$
{\partial f\over \partial g}=
-{3f\over g}{{\cal Q}(x)\over 1+{\cal Q}(x)};
\eqno(A10)
$$
$$
{\partial f\over \partial x}={f\over x}
\left\lbrack 1+
{(1-\alpha\beta)Bx+\alpha {\cal R}(x)\over 1+{\cal R}(x)}
-{(\alpha-1/2){\cal Q}(x)\over 1+{\cal Q}(x)}
\right\rbrack .
\eqno(A11)
$$
The quantities in the above equations are defined in Subsection 2.2.

\section{The Relation between Linear and Virial Radii of Dark Haloes}

As shown by Lahav et al. (1991) the ratio between the virial radius
$r_v$ and the turnaround radius $r_m$ of a dark halo can
approximately be written as 
$$
{r_v\over r_m}={1-\eta/2\over 2-\eta/2},
\eqno(B1)
$$
where 
$$
\eta=2\lambda_0\Omega_0^{-1}\Xi^{-1}(1+z_m)^{-3},
\eqno(B2)
$$
with $z_m$ being the redshift at turnaround and $\Xi$ being
the ratio between the halo and background densities at turnaround.
The ratio between the linear radius ($r_0$) and the virial
radius ($r_v$) of a halo can then be written as
$$
{r_0\over r_v}=(1+z_m)\Xi^{1/3}{2-\eta/2\over 1-\eta/2}.
\eqno(B3)
$$
The value of $\Xi$ depends on $z_m$ as well as on cosmological
parameters. Viana \& Liddle (1996) gave an accurate fitting 
formula for $\Xi$ as a function of $\Omega \equiv \Omega (z_m)$:
$$
\Xi=\left({3\pi\over 4}\right)^2\Omega ^{-f(\Omega)},
\eqno(B4)
$$
where
$$
f(\Omega)=0.76-0.25\Omega \,\,\,\,({\rm open\,\,universe});
\eqno(B5)
$$
$$
f(\Omega)=0.73-0.23\Omega \,\,\,\,({\rm flat\,\,universe}).
\eqno(B6)
$$
Thus, the relation between $r_0$ and $r_v$ is fixed
for a given $z_m$. The value of $z_m$ can be obtained
by the assumption that the age of the universe at the turnaround 
of a halo ($t_m$) is half the age of the universe at its collapse 
($t_c$): $t_m=t_c/2$. For a given redshift of halo collapse,
$z_c$, the time $t_c$ and the redshift $z_m$ can be calculated
from the redshift-time relation. This relation for an open
universe is
$$
t=H_0^{-1}{(1+\Omega_0z)^{1/2}\over (1-\Omega_0)(1+z)}
-{\Omega_0\over 2(1-\Omega_0)^{3/2}}
\ln{(1+\Omega_0z)^{1/2}+(1-\Omega_0)^{1/2}
\over(1+\Omega_0z)^{1/2}-(1-\Omega_0)^{1/2}},
\eqno(B7)
$$
and for a flat universe is
$$
t=H_0^{-1}{2\over 3\lambda_0^{1/2}}
\ln \left\lbrace {\left[\lambda_0(1+z)^{-3}\right]^{1/2}
+\left[\lambda_0(1+z)^{-3}+\Omega_0\right]^{1/2}
\over \Omega_0^{1/2}}\right\rbrace.
\eqno(B8)
$$
\begin{deluxetable}{lcccccccc}
\tablewidth{0pc}
\tablecaption{Simulation Parameters}
\tablehead{
\colhead{Model}
& \colhead{Realizations}
& \colhead{$\Omega_0$} &
\colhead{$\lambda_0$} & \colhead{$\Gamma$} &
\colhead{$\sigma_8$} & \colhead{$L[\mpch]$} 
&\colhead{$N_p$}     & \colhead{$\epsilon[\mpch]$}}
 
\startdata
SCDM0.62 &3   &1.0   &0.0  &0.5   &0.62  &128  &$100^3$ &0.1  \nl
B-SCDM0.62 &5   &1.0   &0.0  &0.5   &0.62  &300  &$128^3$ &0.2  \nl
SCDM1.24 &5   &1.0   &0.0  &0.5   &1.24  &300  &$128^3$ &0.2  \nl
FLAT     &5   &0.2   &0.8  &0.2   &1.00  &128  &$64^3$ &0.1  \nl
OPEN     &5   &0.2   &0.0  &0.2   &1.00  &128  &$64^3$ &0.1  \nl
\enddata
\end{deluxetable}

\begin{deluxetable}{lcccc}
\tablewidth{0pc}
\tablecaption{The Values of $\langle v_1^2\rangle ^{1/2}$, $Q$ and $\gamma$}
\tablehead{
\colhead{Model}
&\colhead{$\langle v_1^2\rangle ^{1/2}$ (model)} 
&\colhead{$\langle v_1^2\rangle ^{1/2}$ (simulation)}
&\colhead{$Q$ } 
&\colhead{$\gamma$ }}

\startdata
SCDM0.62   &713 [762]   &$714\pm 22$   &1.46    &1.74\nl
B-SCDM0.62 &742 [762]   &$694\pm 12$   &1.46    &1.74\nl
SCDM1.24   &1464 [1482] &$1380\pm 33$  &1.55    &1.72\nl
FLAT       &521 [595]   &$524\pm 21$   &1.78    &1.75\nl
OPEN       &571 [642]   &$556\pm 28$   &1.42    &1.77\nl
\enddata
\end{deluxetable}

\begin{figure}
\epsscale{1.0}
\plotone{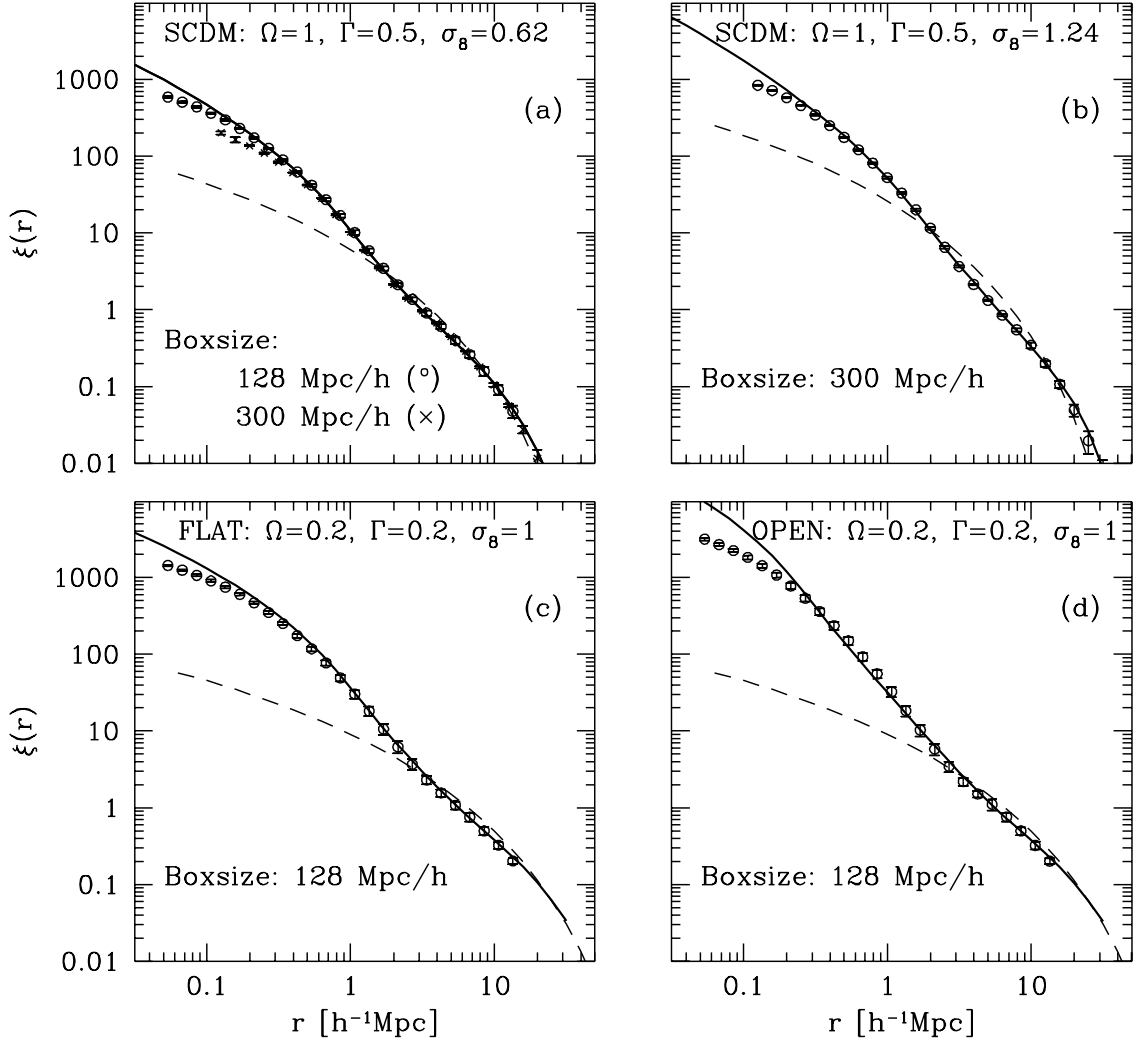}
\caption{
 The two-point correlation functions of dark matter particles
predicted by the analytical model (solid curves) compared
to the results derived from N-body simulations (symbols).
The dashed curves show the two-point correlation functions
given by the linear power spectra. 
The error bars show the scatter among different realizations.
The model parameters and the simulation box sizes
are indicated in each panel.
}\end{figure}

\begin{figure}
\epsscale{1.0}
\plotone{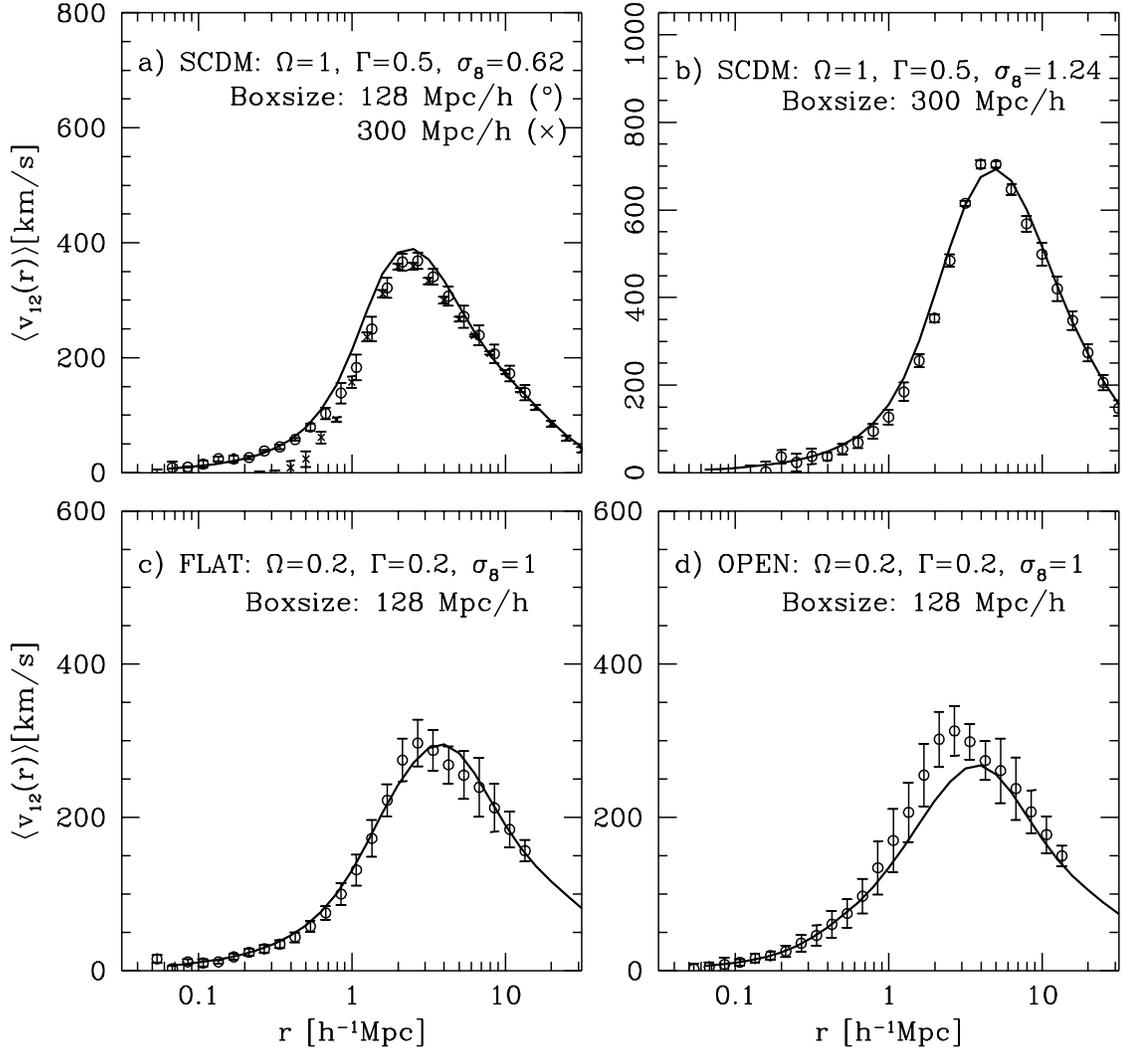}
\caption{
 The average pairwise peculiar velocity of dark matter particles
predicted by the analytical model (solid curves) compared
to the results derived from N-body simulations (symbols).
The error bars show the scatter among different realizations 
}\end{figure}

\begin{figure}
\epsscale{1.0}
\plotone{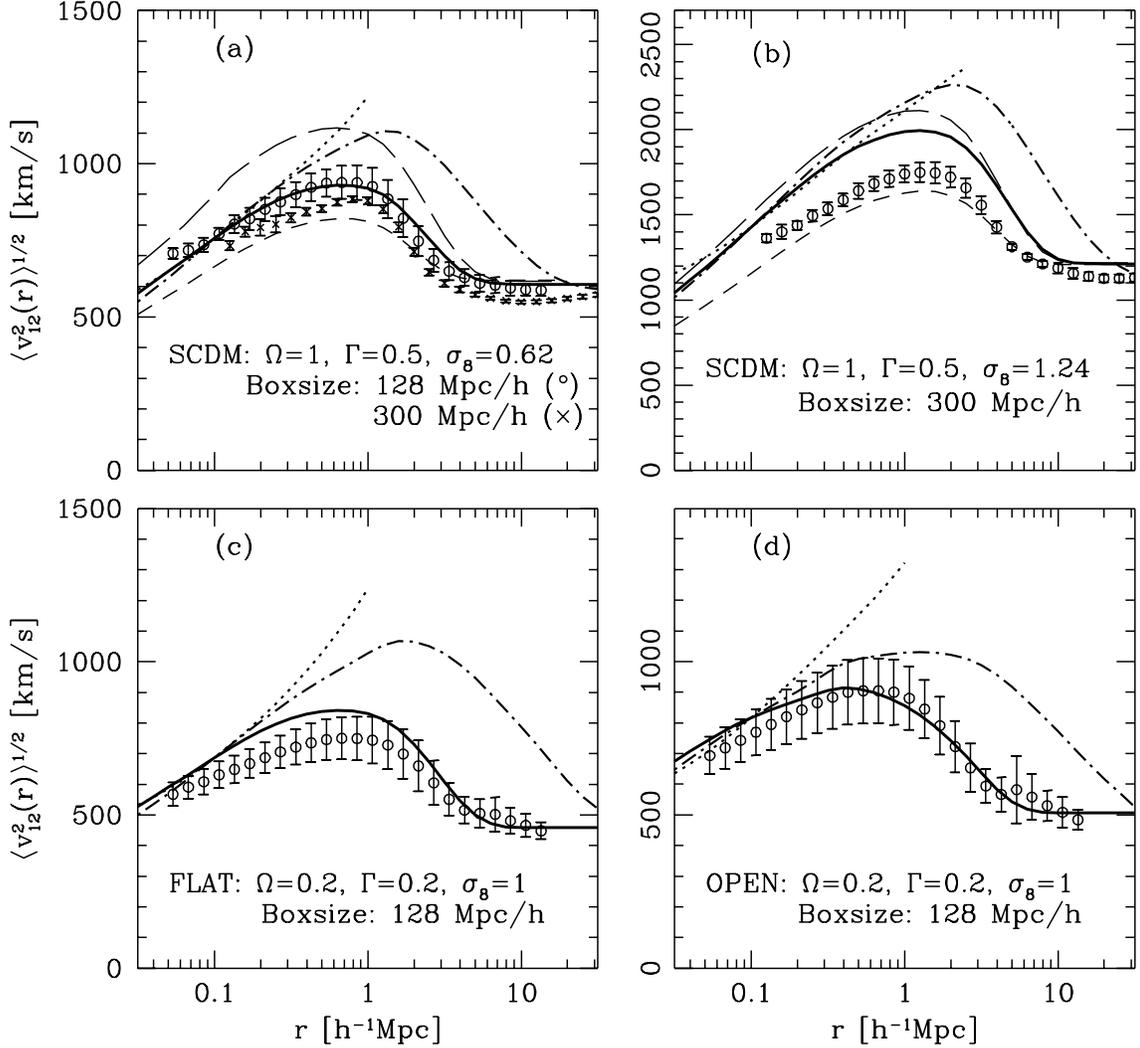}
\caption{
 The pairwise peculiar velocity dispersion of dark matter particles
predicted by the analytical model (curves) at the different stages
of approximation (see Section 2.5.1). The dotted curves
show the predictions of equation (23), while the dot-dashed
and solid curves show the results given by equation (31)
and equation (38), respectively. The two dashed curves
(in a and b) show the predictions of equation (38) with
the concentration factor $c=5$ (lower curve) and 10.
The symbols show results derived from N-body simulations.
Error bars are the scatter among different realizations 
}\end{figure}

\begin{figure}
\epsscale{1.0}
\plotone{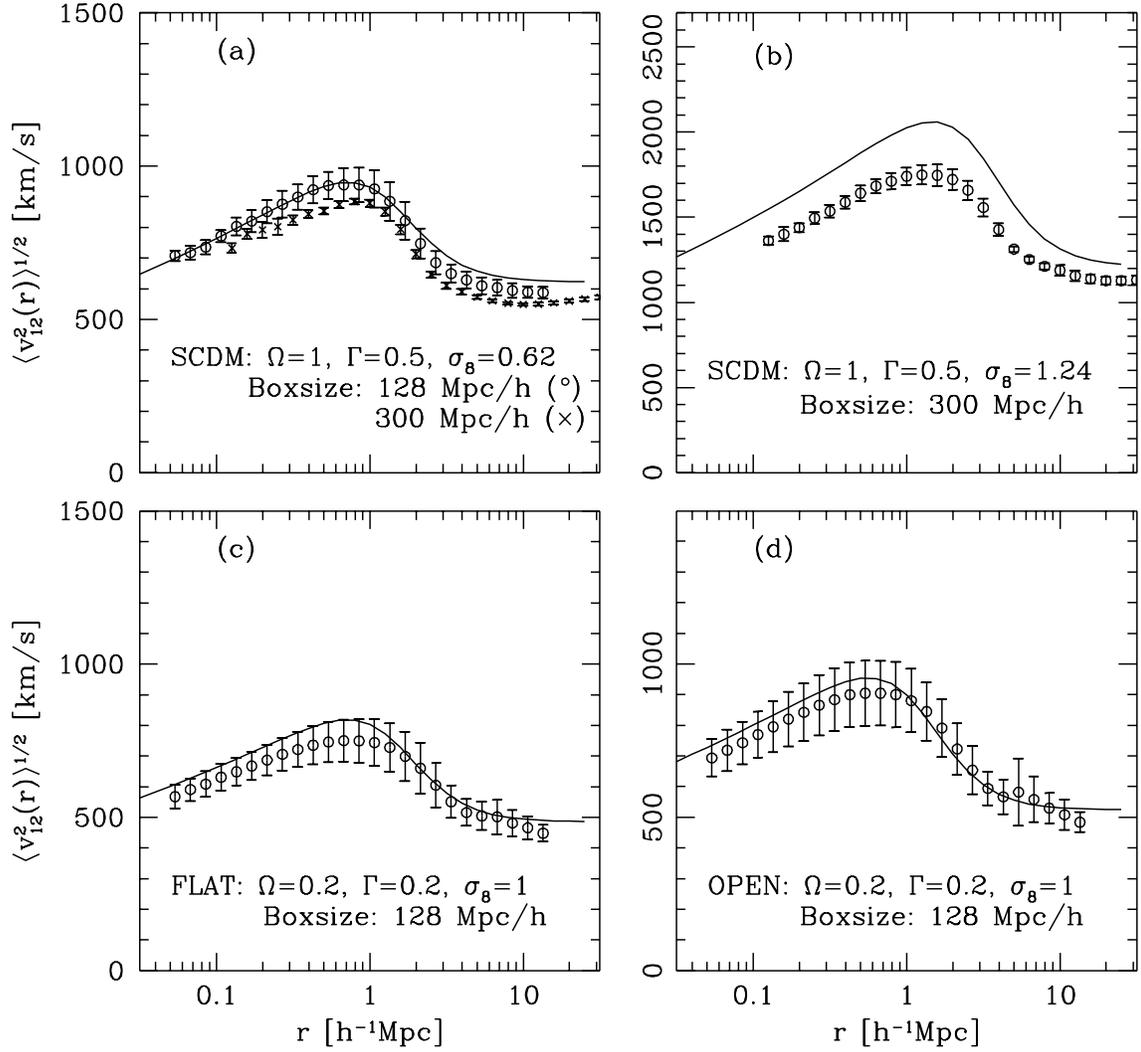}
\caption{
 The pairwise peculiar velocity dispersion of dark matter particles
given by the fitting formula (40) (curves), compared with
the results derived from N-body simulations (symbols).
Error bars are the scatter among different realizations 
}\end{figure}

\begin{figure}
\epsscale{1.0}
\plotone{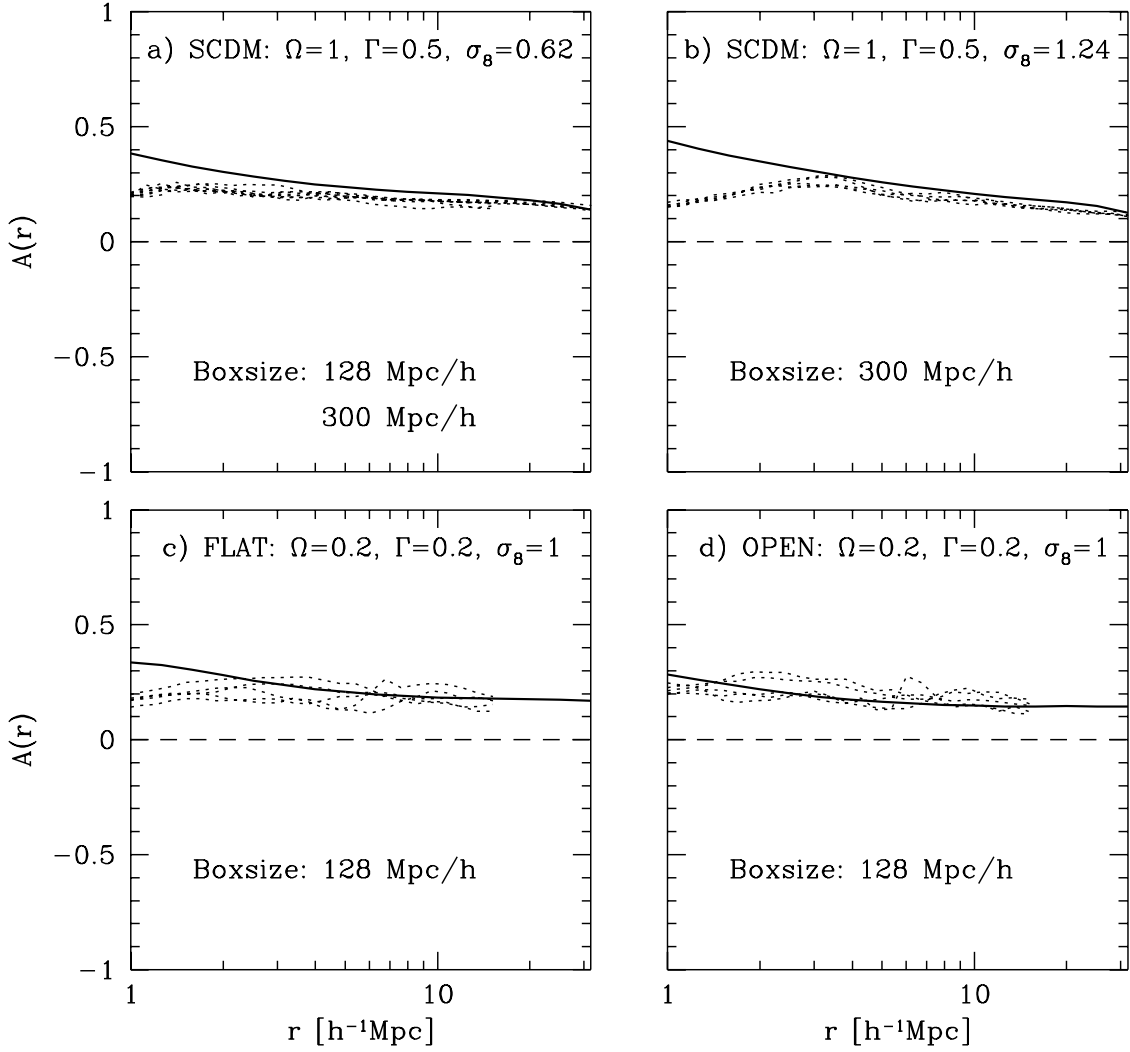}
\caption{
 The velocity-dispersion anisotropy ${\cal A}$
of dark matter particles
predicted by the analytical model (solid curves) 
compared to the results derived from N-body simulations.
Each dotted curve shows the result of simulation in
one particular realization. 
}\end{figure}

\begin{figure}
\epsscale{1.0}
\plotone{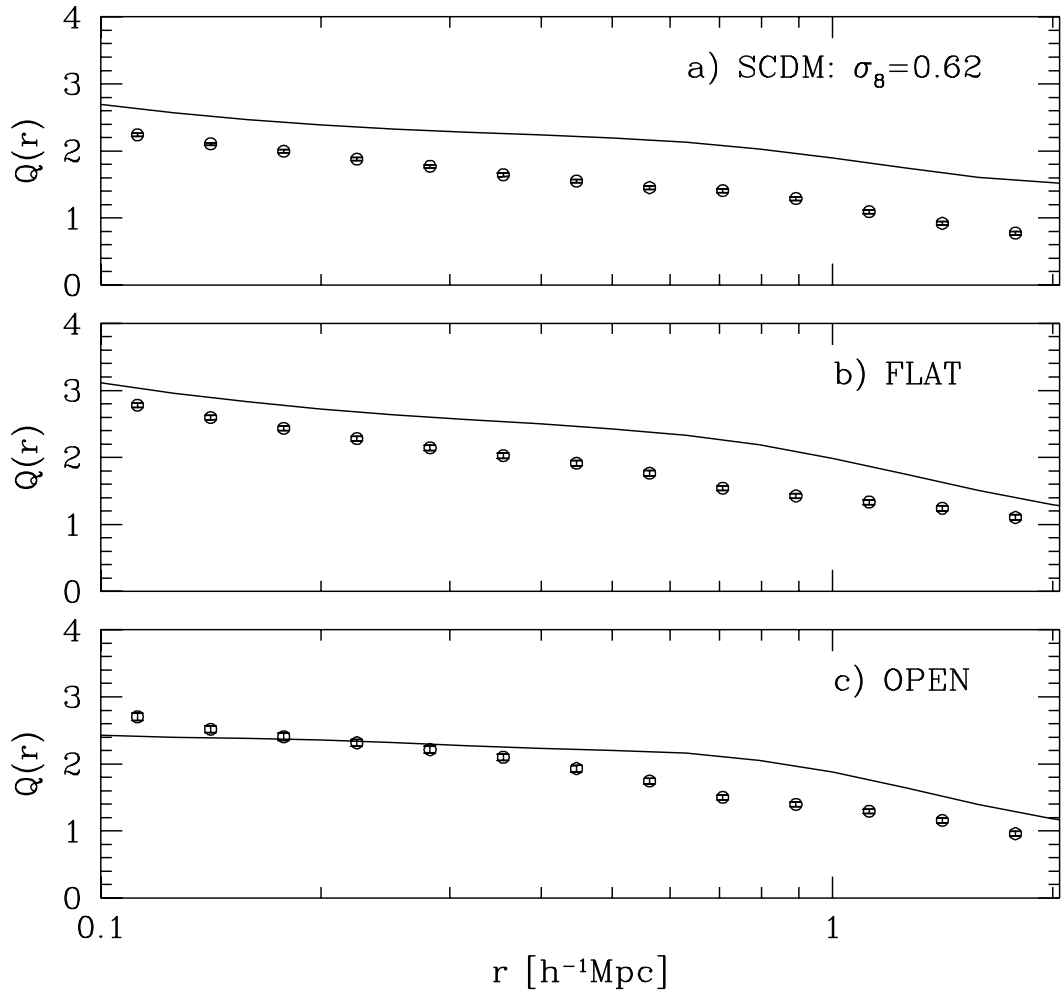}
\caption{
 The amplitude of the three-point correlation function
of dark matter particles
predicted by the analytical model (solid curves) 
compared to the results derived from N-body simulations
(symbols).
The error bars are the scatter among different realizations 
}\end{figure}

\begin{figure}
\epsscale{1.0}
\plotone{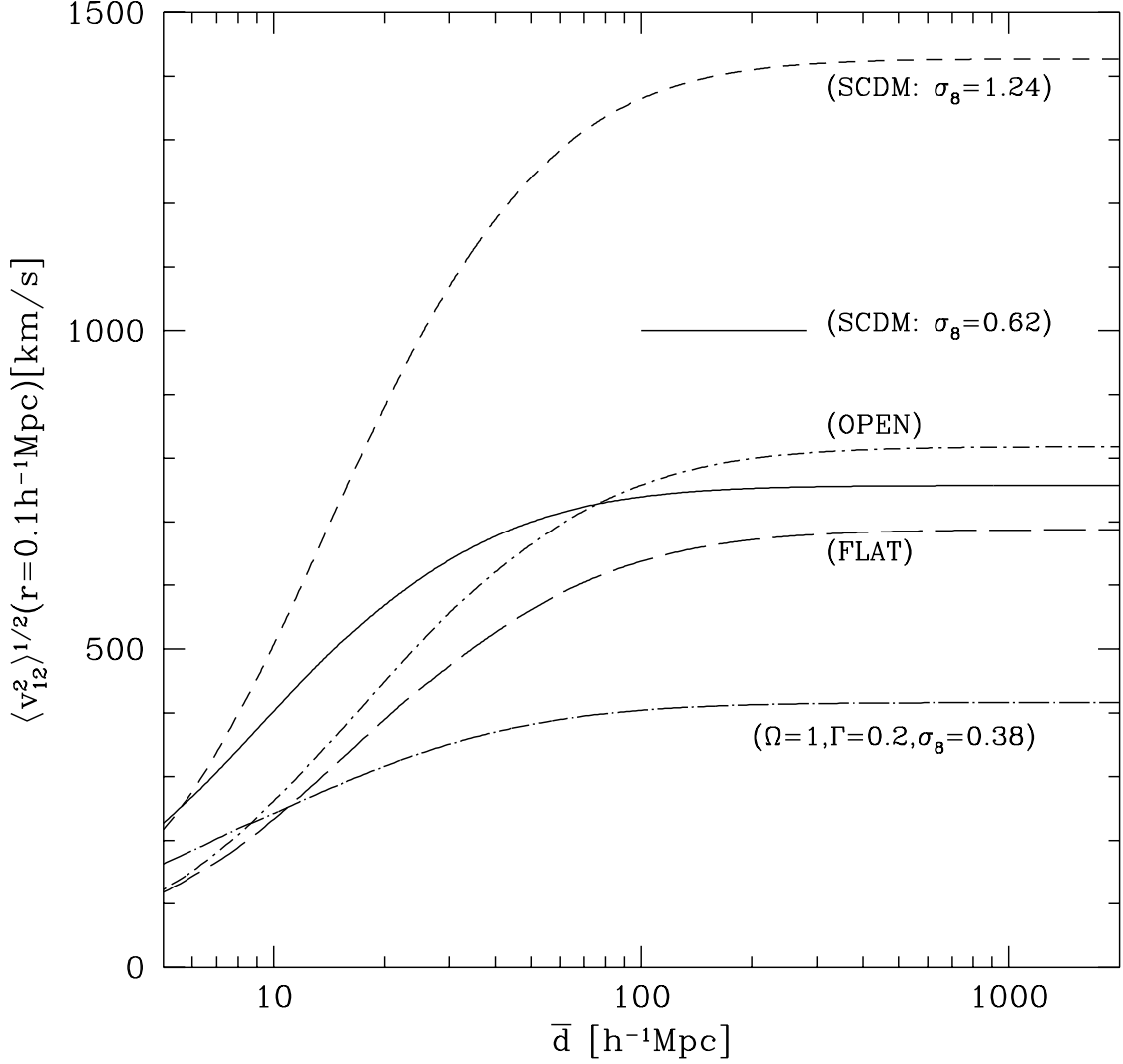}
\caption{
 The pairwise peculiar velocity dispersion 
of dark matter particles at separation $r=0.1\mpch$
as a function of ${\overline d}$, the mean separation of
haloes that are removed before the pairwise peculiar
velocity dispersion is calculated.
Results are shown for the SCDM models ($\Omega_0=1$,
$\Gamma=0.5$) with $\sigma_8=0.62$ and 1.24,
for FLAT and OPEN models ($\Omega_0=0.2$, $\Gamma=0.2$)
with $\sigma_8=1$ and for an additional model
with $\Omega_0=1$, $\Gamma=0.2$ and $\sigma_8=0.38$.
The last three models have the same value of
$\sigma_8\Omega_0^{0.6}$.
}\end{figure}
\end{document}